\documentclass[conference]{IEEEtran}

\ifCLASSINFOpdf
\usepackage[pdftex]{graphicx}
\else
\fi

\usepackage{balance}
\usepackage{cite}
\usepackage{graphicx}
\usepackage{float}
\usepackage{amsmath}
\usepackage[utf8]{inputenc}
\usepackage[final]{pdfpages}
\usepackage{pdfpages}
\usepackage{lipsum}
\usepackage{textcase}
\usepackage{url}
\usepackage{amsmath,esint}
\usepackage{epstopdf}
\usepackage{array}
\usepackage{color}
\usepackage{subcaption}

\usepackage{bm}

\DeclareRobustCommand{\uvec}[1]{{%
  \ifcsname uvec#1\endcsname
     \csname uvec#1\endcsname
   \else
    \bm{\hat{\mathbf{#1}}}%
   \fi
}}

\newcolumntype{P}[1]{>{\centering\arraybackslash}p{#1}}
\newcolumntype{M}[1]{>{\centering\arraybackslash}m{#1}}

\pdfoutput=1

\begin{document}

\title{Indoor Coverage Enhancement for mmWave Systems with Passive Reflectors: Measurements and Ray Tracing Simulations}

\author{
\IEEEauthorblockN{Wahab Khawaja\IEEEauthorrefmark{1}, Ozgur Ozdemir\IEEEauthorrefmark{1}, Yavuz Yapici\IEEEauthorrefmark{1}, Ismail Guvenc\IEEEauthorrefmark{1}, Martins Ezuma\IEEEauthorrefmark{1} and Yuichi Kakishima\IEEEauthorrefmark{2} 
}
\IEEEauthorblockA{\IEEEauthorrefmark{1}Department of Electrical and Computer Engineering, North Carolina State University, Raleigh, NC}
\IEEEauthorblockA{\IEEEauthorrefmark{2}DOCOMO Innovations, Inc., Palo Alto, CA}
Email: \{wkhawaj, oozdemi, yyapici, iguvenc, mcezuma\}@ncsu.edu, kakishima@docomoinnovations.com
	
}

\maketitle
 
\begin{abstract}
The future 5G networks are expected to use millimeter wave~(mmWave) frequency bands, mainly due to the availability of large unused spectrum. However, due to high path loss at mmWave frequencies, coverage of mmWave signals can get severely reduced, especially for non-line-of-sight~(NLOS) scenarios. In this work, we study the use of passive metallic reflectors of different shapes/sizes to improve mmWave signal coverage for indoor NLOS scenarios. Software defined radio 
based mmWave transceiver platforms operating at $28$~GHz are used for indoor measurements. Subsequently, ray tracing~(RT) simulations are carried out in a similar environment using Remcom Wireless InSite software. The cumulative distribution functions of the received signal strength for the RT simulations in the area of interest are observed to be reasonably close with those obtained from the measurements. Our measurements and RT simulations both show that there is significant~(on the order of $20$~dB) power gain obtained with square metallic reflectors, when compared to no reflector scenario for an indoor corridor. We also observe that overall mmWave signal coverage can be improved utilizing reflectors of different shapes and orientations.  


\begin{IEEEkeywords}
Coverage, indoor, mmWave, non-line-of-sight~(NLOS), PXI, ray tracing~(RT), reflector.
\end{IEEEkeywords}

\end{abstract}

\IEEEpeerreviewmaketitle

\section{Introduction}
The use of smart communication devices and the higher data rate applications supported by them have seen a surge in the recent decade. These applications require higher communication bandwidths, whereas the available sub-$6$~GHz spectrum is reaching its limits due to spectrum congestion. With the opening of millimeter wave~(mmWave) spectrum by FCC~\cite{FCC_28G}, various research efforts are underway to use mmWave spectrum for future $5$G communications. However, a major bottleneck for propagation at mmWave frequencies is the high free space attenuation,  especially for the non-line-of-sight~(NLOS) paths. This makes the radio frequency planning very difficult for long distance communications. 

Various solutions to this problem have been proposed in the literature, including, high transmit power, high sensitivity receivers, deployment of multiple access points or repeaters, and beam-forming using multiple antennas. However there are limitations to each of these solutions. Increasing the transmit power beyond a certain level becomes impractical due to regulations, whereas the receiver sensitivity is constrained by the sophisticated and expensive equipment requirement. Similarly, using large number of access points may not be feasible economically. The beam forming requires expensive, complex and power hungry devices, and it may still suffer from NLOS propagation. 

A feasible and economical solution for NLOS mmWave signal coverage is by introducing metallic passive reflectors. This stems from the fact that electromagnetic waves behave similar to light~\cite{Light_EM}. The reflection properties of electromagnetic waves are better at higher frequencies due to smaller skin depth~\cite{reflection} and lower material penetration. Similarly, the diffraction around the edges of reflectors is smaller at mmWave frequencies. These reflectors can act similar to a communication repeater but can operate without electricity and negligible maintenance. Similarly, they have larger life spans, and small initial investment cost when compared with repeaters consisting of active elements. They may even be part of everyday objects, such as street signs, lamp posts, and advertisement boards, that can additionally improve mmWave signal coverage. 

Passive metallic reflectors have been studied and employed in the past for long distance satellite communications~\cite{NASA_refl,Literature4,Literature5}. However, these studies are limited to point-to-point links, whereas, for cellular networks, we may require wide coverage. There are also limited studies available for downlink communications using passive reflectors~\cite{Microwave_refl,Literature6}. This is due to the fact that most of the downlink civilian communications operates at sub-$6$~GHz, where the communication radius is in the kilometer range and few communication repeaters are required. Due to large wavelength, the electromagnetic waves can easily penetrate through most of the building structures without high attenuation, resulting in mostly NLOS communications for the downlink. On the otherhand, mmWave signals observe higher free space path loss and higher penetration loss due to smaller wavelengths. As a result, the communication radius generally shrinks to few hundred of meters. This requires large number of communication repeaters for the downlink and commonly used active repeaters may not be feasible. 

The studies available to date in the literature on using passive reflectors for mmWave coverage enhancement are limited. In~\cite{lit_60GHz_indoor}, indoor coverage analysis at $60$~GHz was carried out due to reflections using simulations. It was observed that at $60$~GHz, the coverage in the NLOS was dependent solely on the reflections. A parabolic passive reflector is used for outdoor coverage enhancement at mmWave frequencies in~\cite{Literature3}, that reflects incoming signal power from base station to users in the building shadowed zones. Numerical results indicate better coverage in the shadowed zones using reflectors. In \cite{Literature1}, a parabolic reflector is used behind a patch antenna operating at $60$~GHz of a hand held device. A gain of $19$~dB - $25$~dB is reported after the introduction of parabolic reflector that helps to counter the finger shadowing while operating the device. In~\cite{Literature2}, reflecting properties of different building materials both in the indoor and outdoor environments were calculated using channel measurements at $60$~GHz.  

\begin{figure*}[!t]
	\begin{subfigure}{0.33\textwidth}
	\centering
	\includegraphics[width=\textwidth]{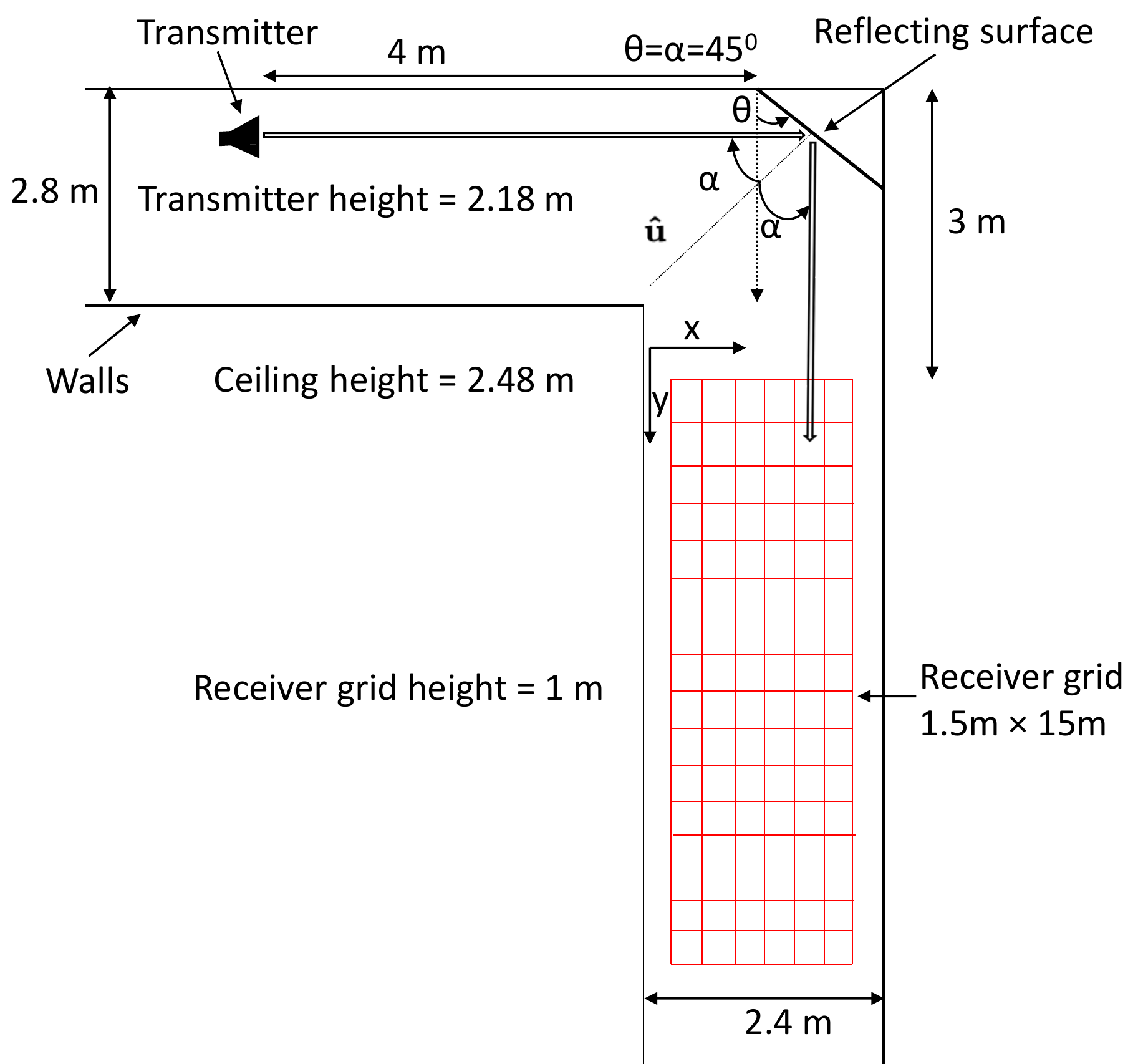}
	\caption{}
    \end{subfigure}			
	\begin{subfigure}{0.33\textwidth}
	\centering
    \includegraphics[width=\textwidth]{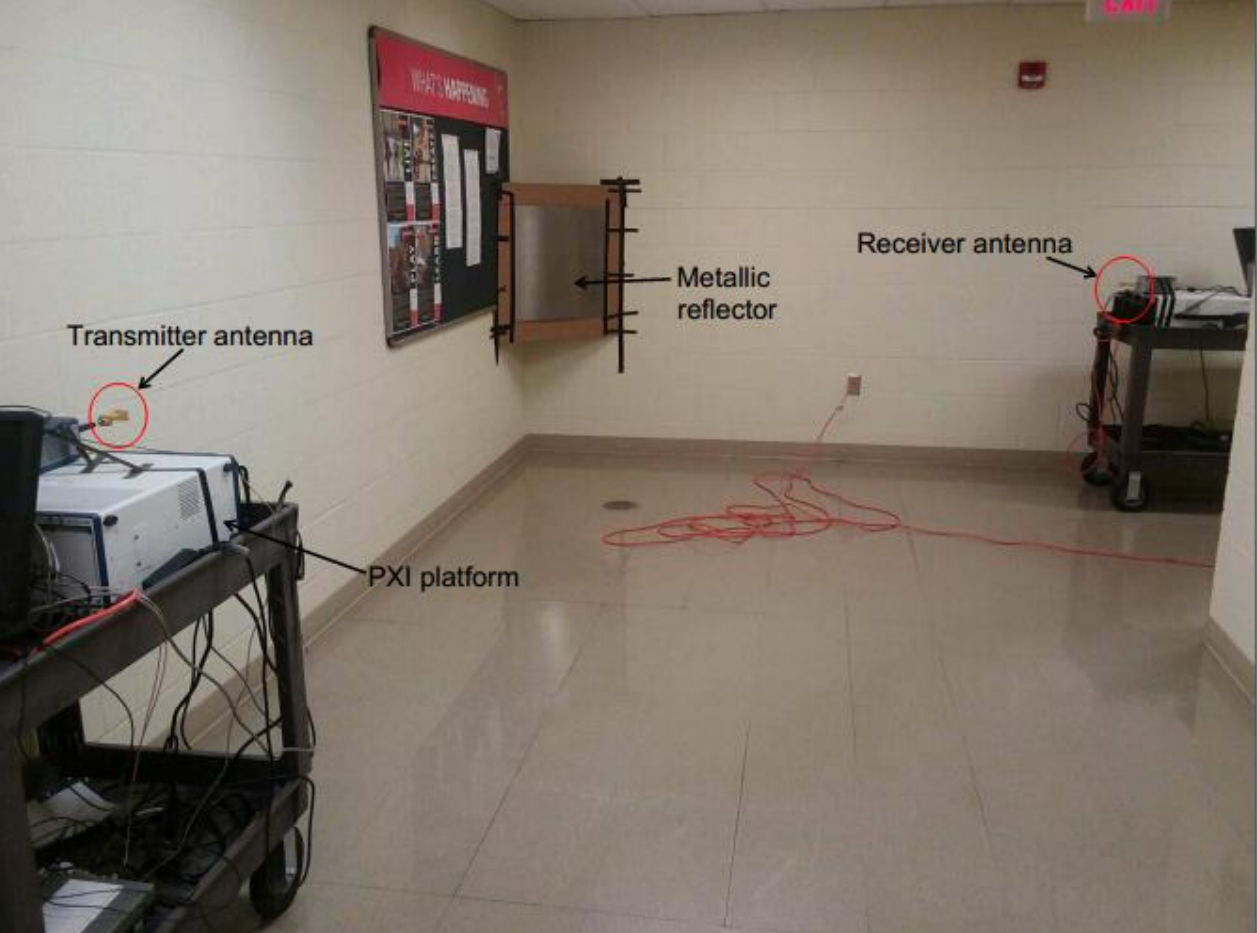}
	 \caption{}
     \end{subfigure}
     \begin{subfigure}{0.33\textwidth}
	\centering
    \includegraphics[width=\textwidth]{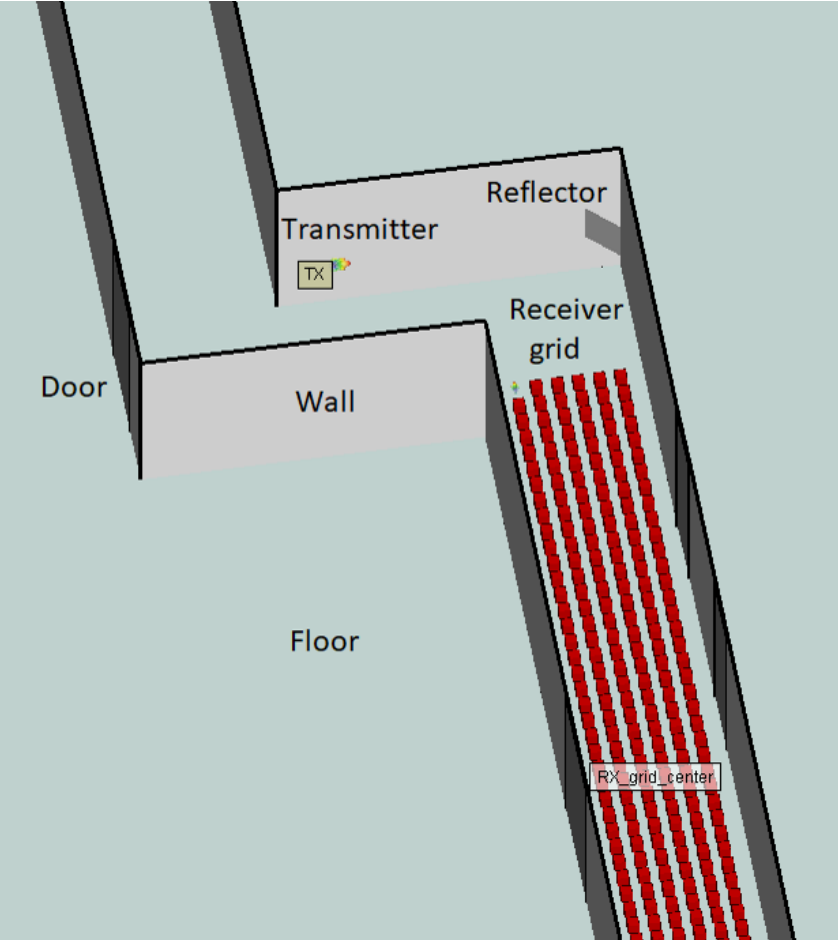}
	 \caption{}
     \end{subfigure}
     \vspace{-1mm}
    \caption{(a) Geometrical model in the azimuth plane for indoor scenario with a reflecting surface deployed at the corner of a corridor; (b) Measurement scenario in the basement corridor of Engineering Building~II at North Carolina State University for flat square sheet aluminum reflector $24\times24$~in$^2$ at an azimuth angle, $\theta = 45^{\circ}$; (c) Simulations of the measurement scenario in Wireless InSite.}\label{Fig:scenario}\vspace{-5mm}
\end{figure*}

To the best of our knowledge, there are no empirical studies available in the literature, on the use of metallic reflectors for downlink coverage enhancement at $28$~GHz. In this work, we have performed measurements for indoor NLOS mmWave propagation scenario at $28$~GHz using National Instruments PXI platform employing different shape and size of metallic reflectors using the setup shown in Fig.~\ref{Fig:scenario}. The received power is measured over an NLOS grid in an indoor corridor. We observe more directional power distribution for flat square sheet reflectors as compared to sphere and cylinder reflectors. Furthermore, we can steer effectively the incident power on a flat sheet reflector to a given direction by changing the corresponding azimuth and elevation angles. With $24\times24$ in$^2$ and $33\times33$~in$^2$ flat square sheet metallic reflectors, a median gain of $20$~dB was observed as compared to no reflector case. The measurement results are compared with the simulation results obtained using Wireless InSite ray tracing~(RT) software by creating a similar indoor environment and incorporating diffuse scattering phenomenon in the simulations. 

\section{Received Power/Coverage Enhancement} 
As mentioned above, electromagnetic waves exhibit similar behavior as light. One of the characteristic feature of light is reflection. In this section, we will discuss about the received power enhancement in NLOS areas using metallic reflectors. 

\subsection{Shape of Reflectors} \label{Section:Shapes}
Shapes of reflectors can vary dependent on the application: e.g., well known primary reflectors used at the backside of the antenna are usually designed to focus the energy beam in a given direction and are usually parabolic in shape. Whereas, for the secondary reflectors, shapes may vary. In the past, large flat metallic sheet reflectors are used for point to point communications among the base stations. However, for down link communications, different shapes of the reflectors can be considered dependent on the coverage requirements. In the rest of the paper, we will focus on the downlink mmWave communication links employing secondary reflectors.   

Metallic flat sheet reflectors have higher cross section area exposed to the incident energy as compared to the cylinder and sphere reflectors. Additionally, the radar cross section~(RCS) pattern is very directive. Therefore, we observe higher amount of energy reflected from large exposed area in a given direction, depending on the orientation of the reflector. For curved shaped reflectors including cylinder and sphere, the reflection characteristics are also dependent on the cross section area. The curved reflectors can either converge or diverge the incoming rays dependent on the side exposed to the incoming beam. Other complex shape reflectors such as sawtooth reflectors can be employed for obtaining different scattering patterns. 


\subsection{Size of Reflectors} \label{Section:Sizes}
The ideal size of the reflector is an important design consideration and it is related to the radiation pattern of the transmitted energy, distance, and location of the reflector from the transmitter for a given coverage area. Considering a directional transmit source with respective half-power radiation beamwidths at the elevation and azimuth planes, the transmitted directional beam can be represented as occupying a certain part of a sphere. The radius of sphere and surface area of the transmitted radiation beam on the sphere is dependent on the distance from the source. For simplicity, we can approximate the spherical surface of the transmitted radiation beam on the sphere as a rectangle\cite{Antenna_approx}. It can be observed that as the radius doubles, the energy from the source spreads at four times the surface area, resulting in one fourth of intensity as compared to the source. Therefore, in order to reflect maximum incident energy at a given distance from the source, the size of the flat reflector should be comparable to the surface area extended by the beam at that distance. On the other hand, from a communication operator's point of view, the size of reflector should be as small as possible considering construction, deployment costs and regulations~(e.g. strong winds, earthquakes). 

\subsection{Energy Steering}\label{Section:Energy_Steering}
In order to dynamically provide coverage to NLOS areas, reflectors can be manually steered similar to beam steering. Depending on the coverage area, the reflector can be oriented in the azimuth and elevation planes accordingly. In order to estimate the reflected ray direction in space, law of reflection may be used i.e., for an incident ray on the reflector at a given surface normal, the reflected ray will also have the same angle as that of the incident ray with the surface normal.

The energy steering mechanism can be explained as follows: If we consider parallel set of straight rays incident on the two edges of the reflector whose surface normal is oriented at angle $\alpha$ with respect to the incident beam, then we observe two parallel reflected rays at angles $2\alpha$ with respect to the incident beam. This angular region has a width proportional to the width of the reflector. 
If the majority of the incident beam is bounded within this width, we expect that the reflected beam will have the same width at an angle $2\alpha$. 

Overall, in order to steer maximum energy, we need a very narrow transmitted beam, especially for long distances. If the distance between the transmitter and the reflector is very large, we will require large size reflectors, comparable to the area of the beam at that point. The steering angle $\theta$ cannot be too large as the effective area given by $A_{\rm e} = A_{\rm Refl}\cos \alpha$, exposed to the incident beam will reduce, where $A_{\rm e}$, and $A_{\rm Refl}$ are the effective and actual reflector areas.   


\section{Propagation Measurements at $28$ GHz Indoor}
Measurements were carried out in the basement corridor of Engineering Building~II at North Carolina State University. The geometrical setup, and the corresponding measurement and simulation environment for indoor channel measurements are shown in Fig.~\ref{Fig:scenario}. The receiver is moved at different positions in the $(x,y)$ plane of the corridor to form a receiver grid. The size of the $(x,y)$ receiver grid is $(1.5\rm{m},15\rm{m})$ such that each measurement block is $0.3$m$\times0.3$m. A similar geometry is generated using  the Remcom Wireless InSite RT software to compare with the measurement outcomes and will be explained in Section~\ref{Sec:RayTracing}.

\begin{figure*}[!t]
\centering

	\begin{subfigure}{.6\textwidth}
	\centering
	\includegraphics[width=\textwidth]{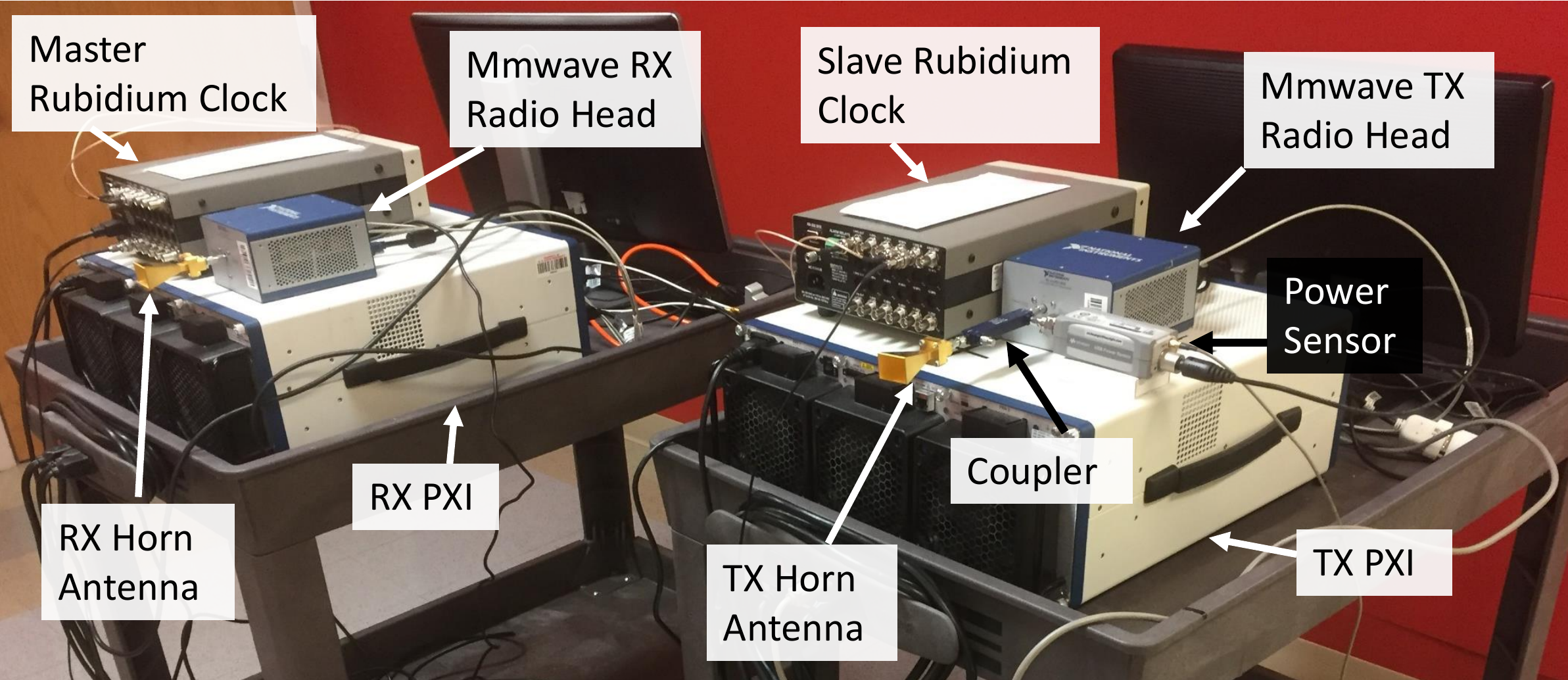}
	\caption{\smallskip}
    \end{subfigure}
    
	\begin{subfigure}{0.2\textwidth}
	\centering
    \includegraphics[width=\textwidth]{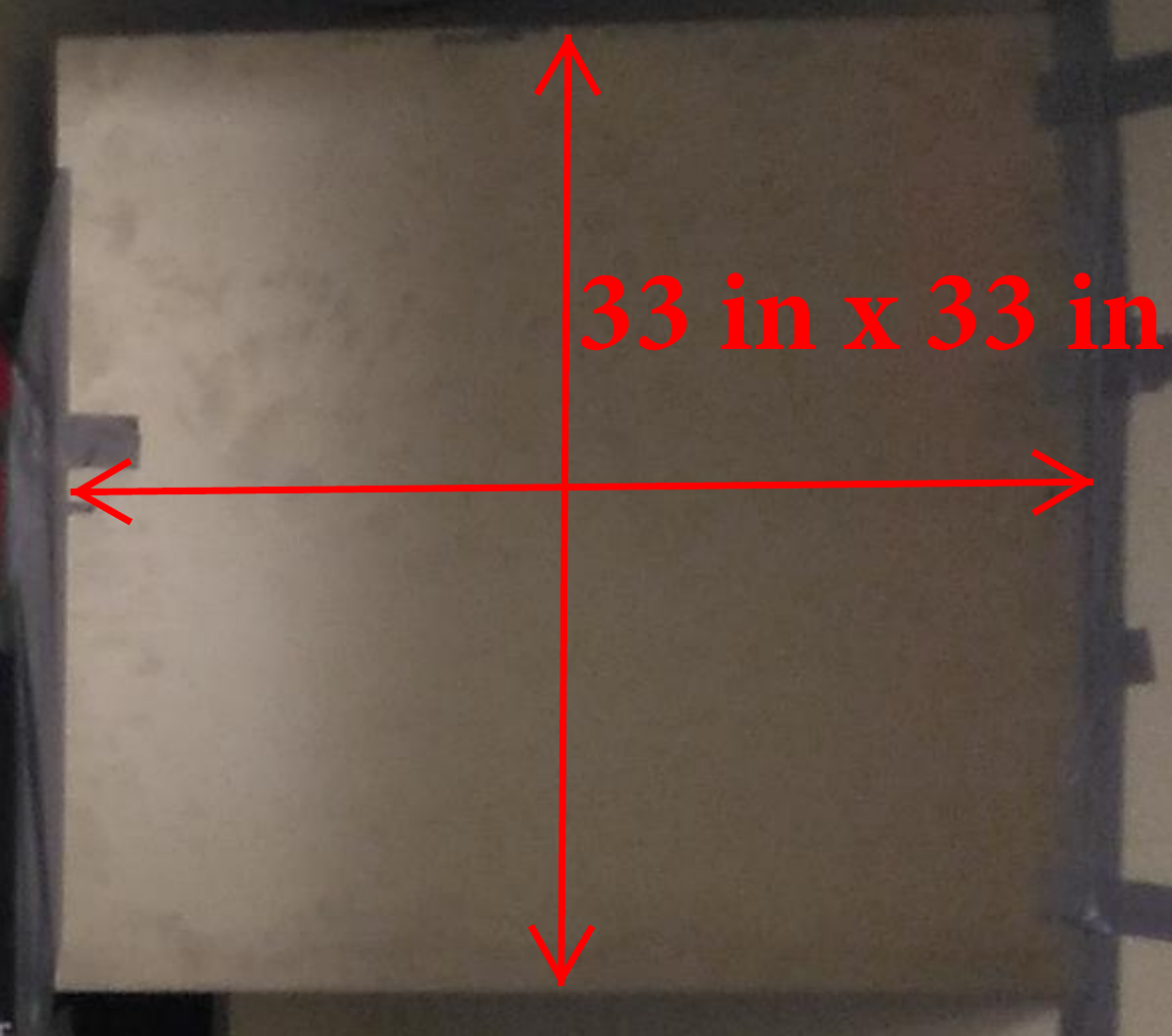}
	 \caption{} 
     \end{subfigure}
     \begin{subfigure}{0.12\textwidth}
	\centering
    \includegraphics[width=\textwidth]{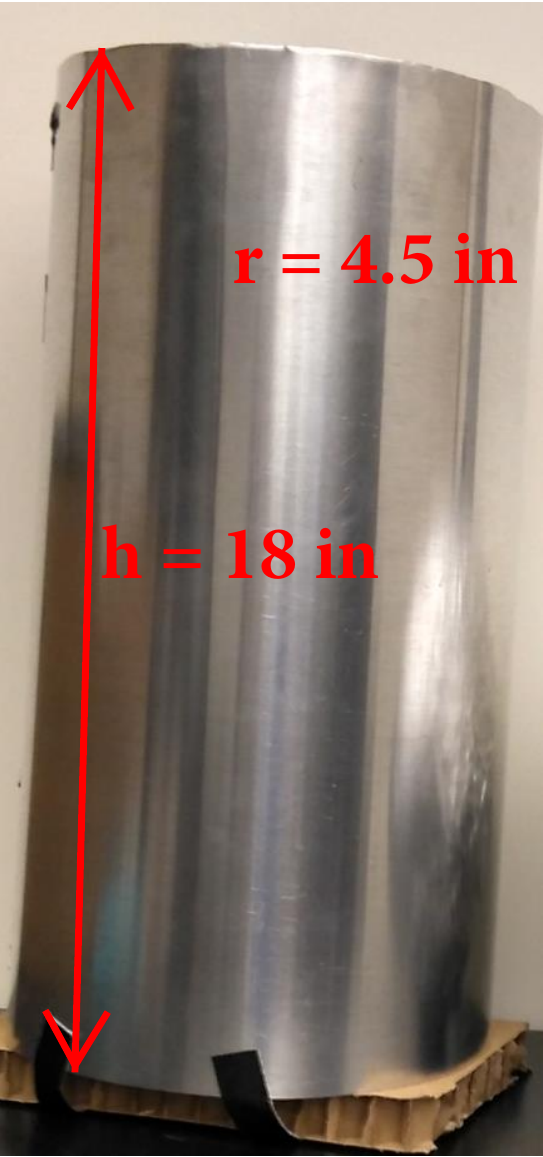}
	 \caption{}
     \end{subfigure}
     \begin{subfigure}{0.18\textwidth}
	\centering
    \includegraphics[width=\textwidth]{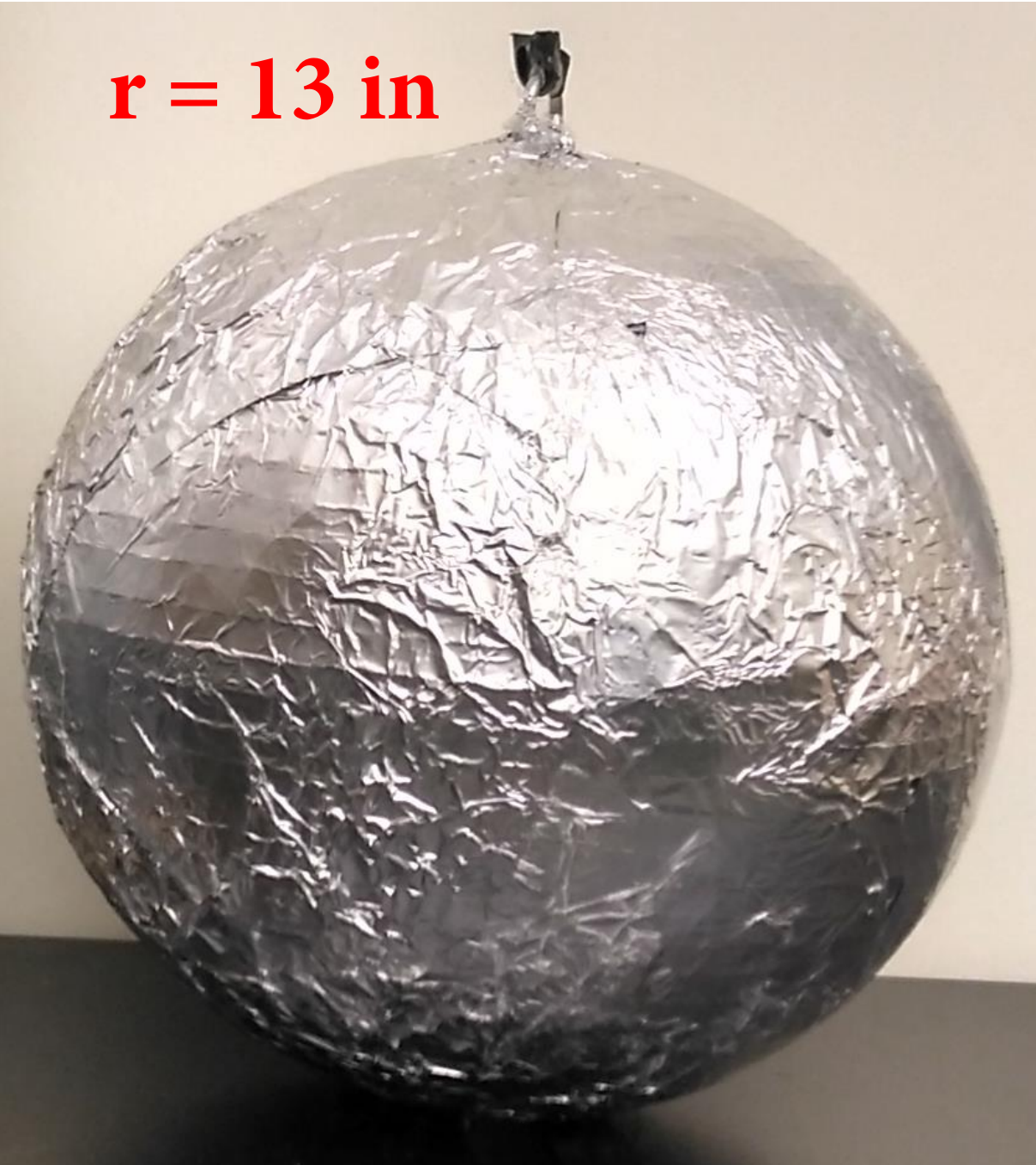}
	 \caption{}
     \end{subfigure}
     \begin{subfigure}{0.18\textwidth}
	\centering
    \includegraphics[width=\textwidth]{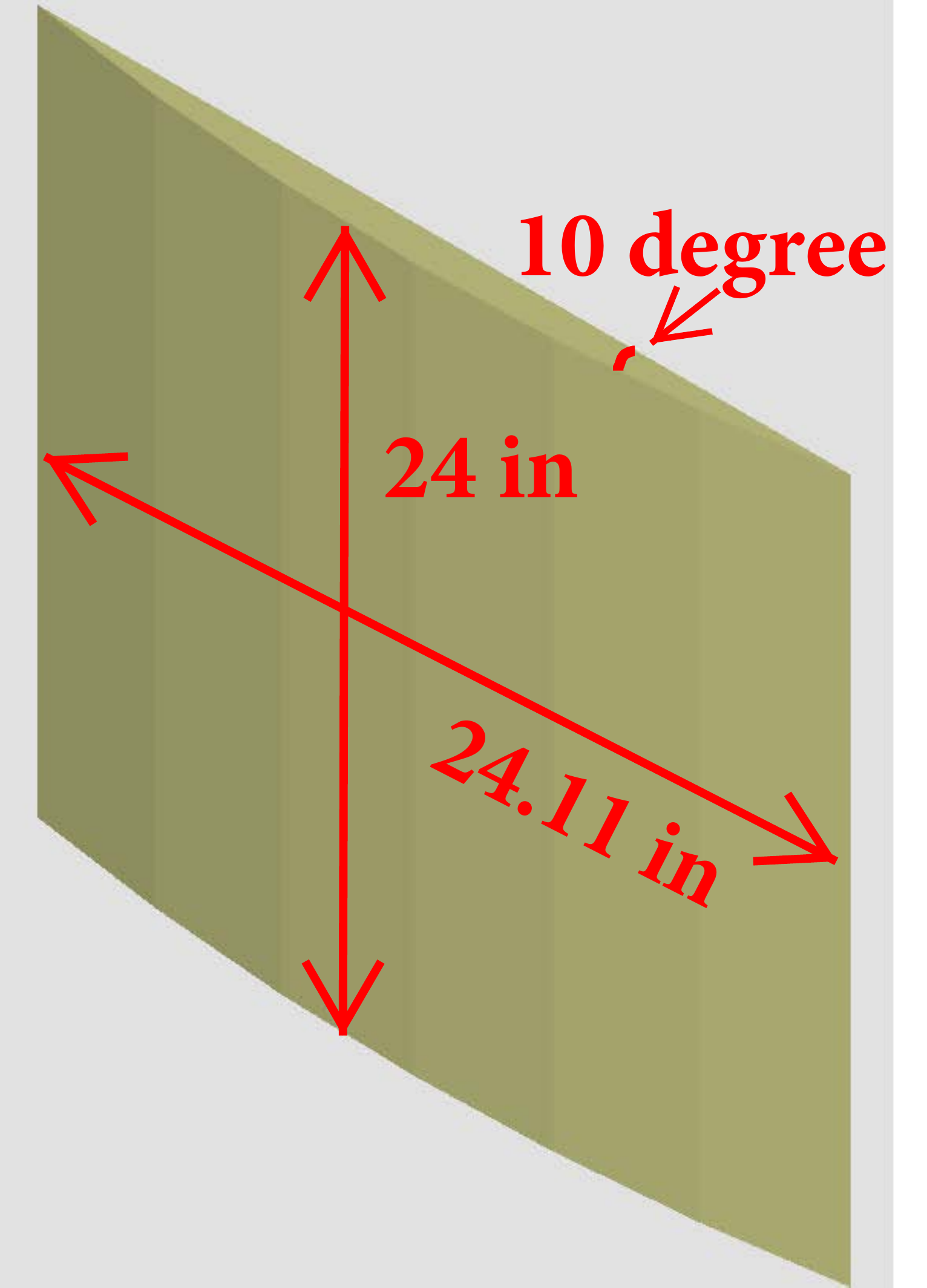}
	 \caption{}
     \end{subfigure}
     \vspace{-1mm}
    \caption{(a) Channel sounder setup, (b) $33\times33$ in$^2$ flat reflector, (c) cylinder reflector, (d) sphere reflector, (e) Curved shape reflector (at 10~degrees of curve). The curved shape reflector is used only in the simulations, while all other reflector shapes are used in both measurements and simulations.}\label{Fig:setup_reflectors}\vspace{-5mm}
\end{figure*}

The measurements were performed using NI mmWave transceiver system at $28$~GHz~\cite{NImmwave} as shown in Fig.~\ref{Fig:setup_reflectors}(a). The system consists of two PXI platforms: one transmitter and one receiver. There are two rubidium~(Rb) clocks used at the transmitter and the receiver sides that provide common $10$~MHz clock and pulse per second (PPS) signal. The output from the PXI intermediate frequency (IF) module is connected to the mmWave transmitter radio head that converts the IF to $28$~GHz. Similarly, at the receiver side, the mmWave radio head down converts $28$~GHz RF signal to IF.

The digital to analog converter at the transmitter and the analog to digital converter at the receiver have a sampling rate of $3.072$~GS/s. The channel sounder supports $1$~GHz and $2$~GHz modes of operation. The measurements for this paper are performed using the $2$~GHz mode where the sounding signal duration is $1.33$~$\mu$s, which also is the maximum measurable excess delay of the sounder. This mode provides a $0.65$~ns delay resolution in the delay domain, corresponding to $20$~cm distance resolution. The analog to digital converter has around $60$~dB dynamic range and this system can measure path loss up to $185$~dB. The transmit power for the experiment is set to $0$~dBm. A power sensor measures the power at the output of the mmWave transmitter front end using an RF coupler. The power sensor lets us convert measurements in dB units into dBm units.

In order get accurate channel measurements, we need to characterize the non-flat frequency response of the measurement hardware itself, and subsequently do a calibration to compensate for the impulse response due to the hardware. For calibration purposes a cable with fixed attenuators connects the transmitter to the receiver. Assuming the cable and the attenuators have flat response, the channel response of the hardware is measured. During actual measurements, the hardware response is equalized assuming hardware response does not vary over time. After this equalization we obtain the response of the actual over the air channel.

The antennas that are used at the transmitter and the receiver are linearly polarized pyramidal horn antennas \cite{Horn_antenna_sage}, having a gain of $17$~dBi and half power beam-widths of $26$~ and $24$ degrees in the E and H planes, respectively.  

To improve the coverage area in NLOS receiver region in the corridors, we use aluminum flat square sheet reflectors with different sizes, a cylinder, and, a sphere as shown in Fig.~\ref{Fig:setup_reflectors}(b), Fig.~\ref{Fig:setup_reflectors}(c), and Fig.~\ref{Fig:setup_reflectors}(d). These reflectors are placed at the corner of the walls facing the corridor as shown in~Fig.~\ref{Fig:scenario}. The aluminum sheet used is 5086-H32 having a thickness of $.063$~in. Three flat square sheets with side lengths of $12$~in, $24$~in, and $33$~in respectively, are used in the measurements. A metallic cylinder of radius $4.2$~in and height $18$~in is used, whereas a mirror ball covered with aluminum sheet having a diameter of $13.5$~in is used. The surface areas of $24\times24$~in$^2$, flat reflector, cylinder, sphere have similar cross sectional area. 

In order to place different flat reflectors at the same plane, a cardboard of size $33\times33$~in$^2$ is used as a reference as shown in Fig.~\ref{Fig:scenario}(b). The center of the cardboard is aligned to the center of the bore-sight axis of the antenna. Different sized reflectors are placed such that their centers are aligned to the center of the cardboard. Similarly, the bore-sight axis of the antenna is aligned to the center of the cylinder and sphere. There is no orientation of the reflectors in the vertical plane.

\section{Ray Tracing Simulations at 28 GHz}\label{Sec:RayTracing}
Simulations for the passive metallic reflectors at mmWave frequencies are performed using Remcom Wireless InSite RT software, replicating the indoor experimental environment as shown in Fig.~\ref{Fig:scenario}(c). The red blocks in the figure represent the individual receiver points in the grid. A sinusoidal sounding signal at $28$~GHz is used, and the transmit power is set to $0$~dBm. Horn antennas~\cite{Horn_antenna_sage}, similar to used in the measurements, are used at both transmitter and the receiver grid. 

In addition to specular reflection at mmWave frequencies, diffuse scattering also occurs dominantly due to comparable size of wavelength of the transmitted wave and the dimensions of the irregularities of the surfaces that it encounters. In the simulations, diffuse scattering feature has been used to take into account this factor. The diffuse scattering model used in the simulations is directive model. Only the diffuse scattering coefficient is changed for different materials, whereas the other model parameters remain the same. Diffuse scattering coefficient of different materials used in the simulations are $0.1$, $0.2$, $0.25$ and $0.3$, for perfect conductor, concrete, ceiling board, and layered dry wall, respectively. The materials with higher roughness are assigned higher diffuse scattering coefficient.

The received power is obtained and summed non-coherently from the received multipath components~(MPCs) at a given receiver location. This does not involve the phase of each MPC to be considered in the received power calculation. The similar is done for measurements.  

The walls, floor, ceiling, door and reflector materials are selected such that they are similar to the actual measurement environment setup as much as possible. The \emph{ITU three layered drywall} is used for walls and ITU \emph{ceiling board} is used for ceilings, \emph{concrete} is used for floor, and a \emph{perfect conductor} is used for the door and the metallic reflector. All the materials are frequency sensitive at $28$~GHz. The dimensions of the simulation setup are the same as in Fig.~\ref{Fig:scenario}(a).

\begin{figure*}[!t]
	\begin{subfigure}{0.5\textwidth}
	\centering
	\includegraphics[width=\columnwidth]{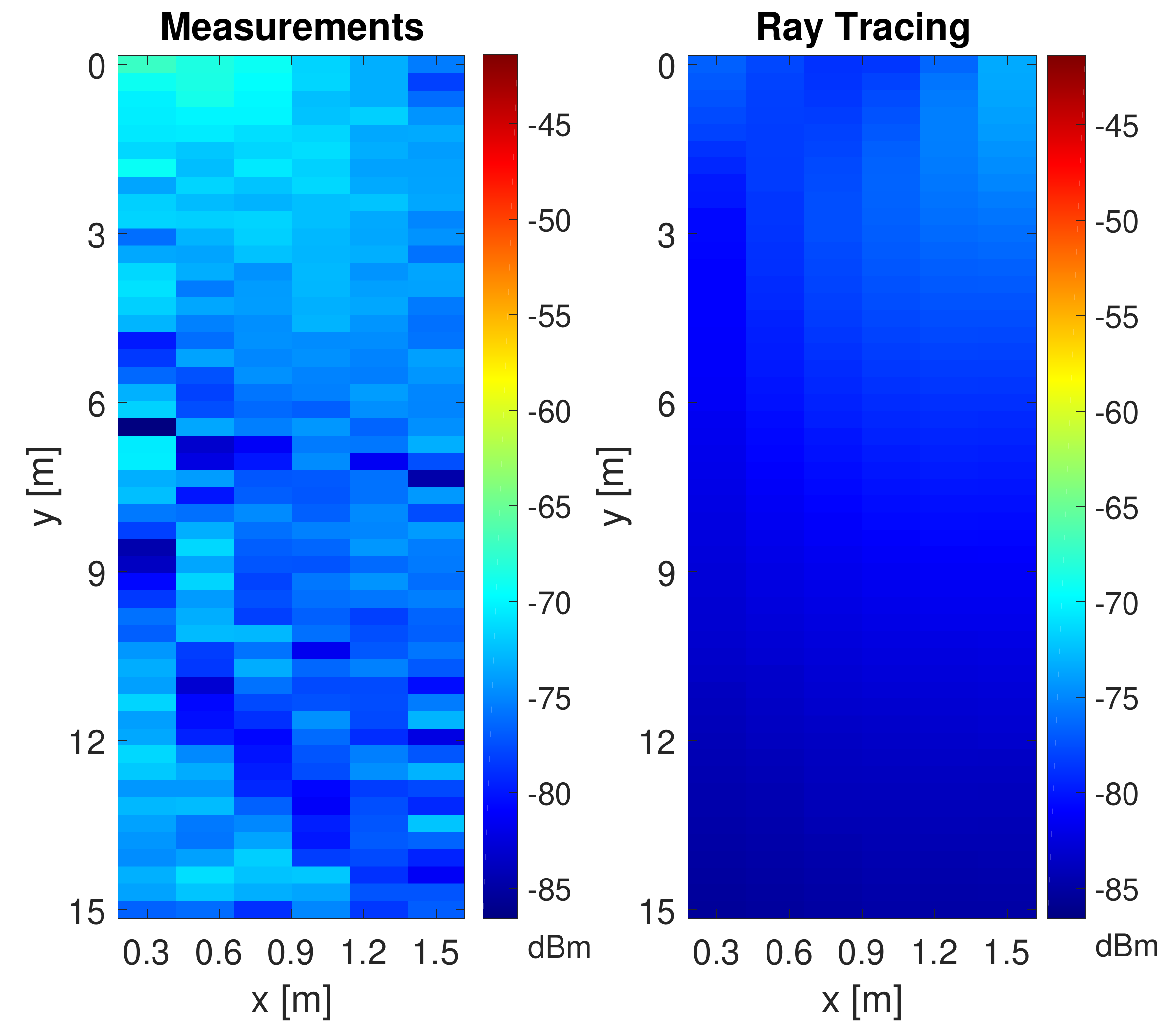}
	\caption{}
    \end{subfigure}			
	\begin{subfigure}{0.5\textwidth}
	\centering
    \includegraphics[width=\columnwidth]{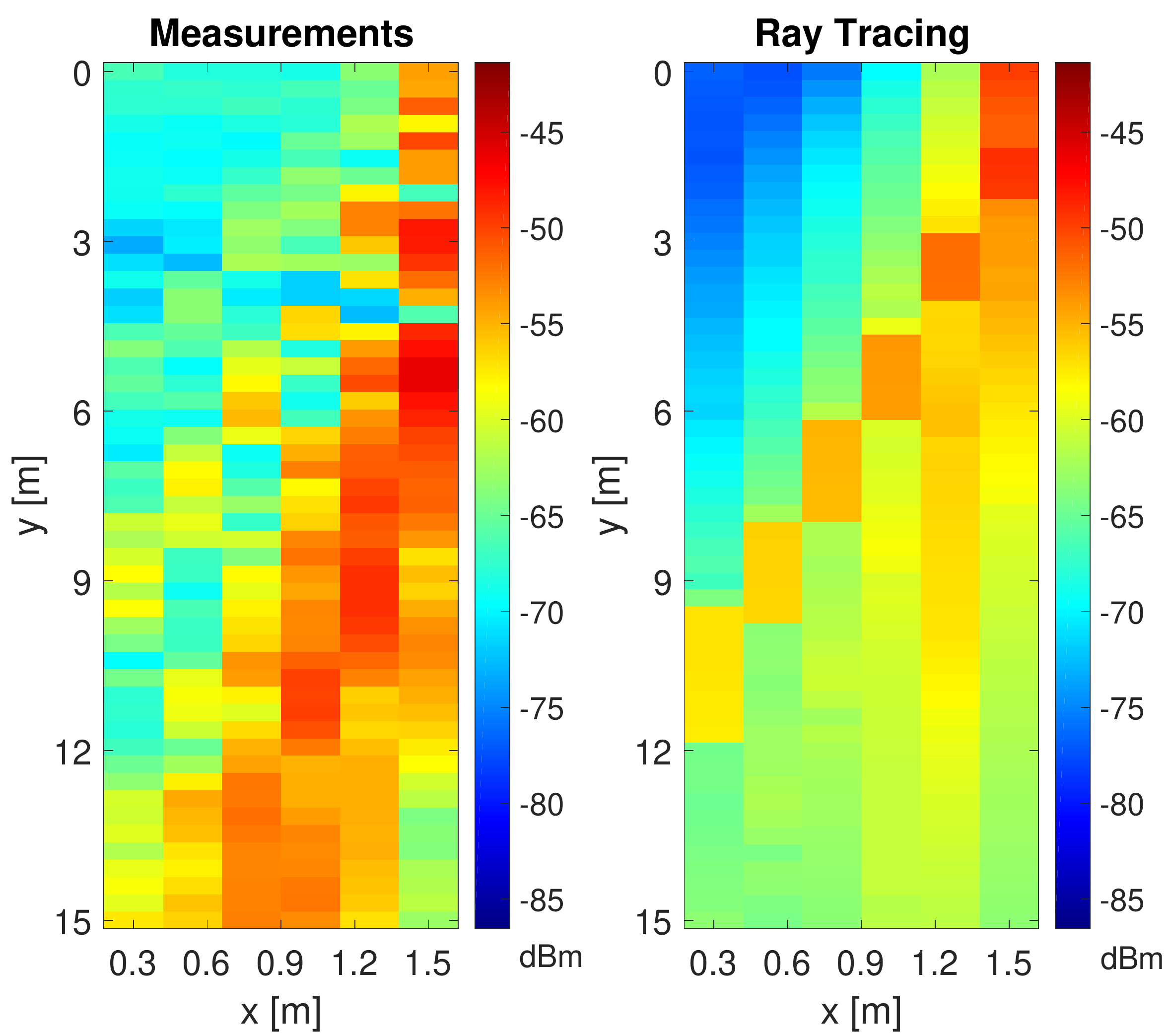}
	 \caption{}
     \end{subfigure}
    	 	
	\begin{subfigure}{0.5\textwidth}
	\centering
    \includegraphics[width=\columnwidth]{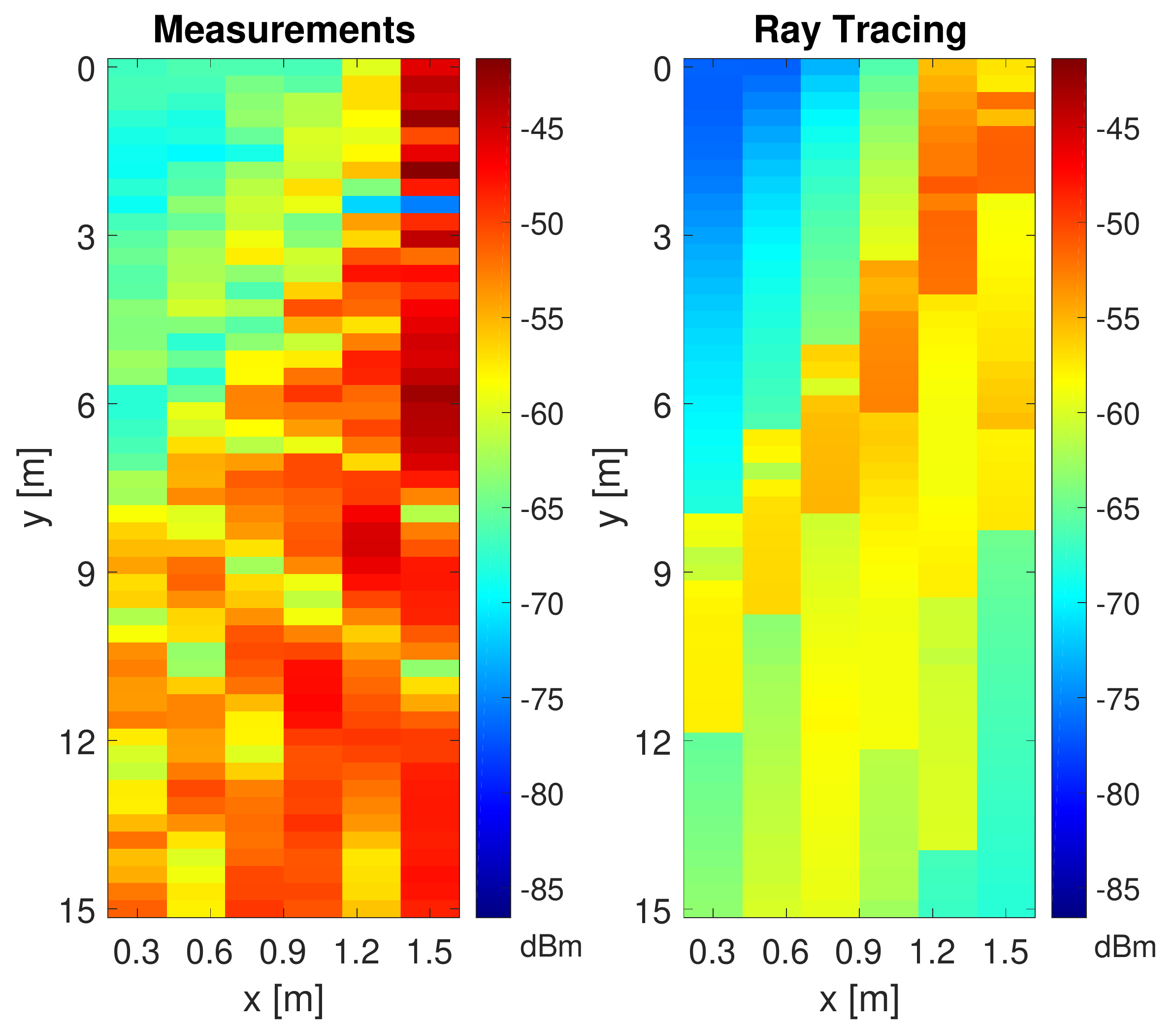}
	 \caption{}
     \end{subfigure}
     \begin{subfigure}{0.5\textwidth}
	\centering
    \includegraphics[width=\columnwidth]{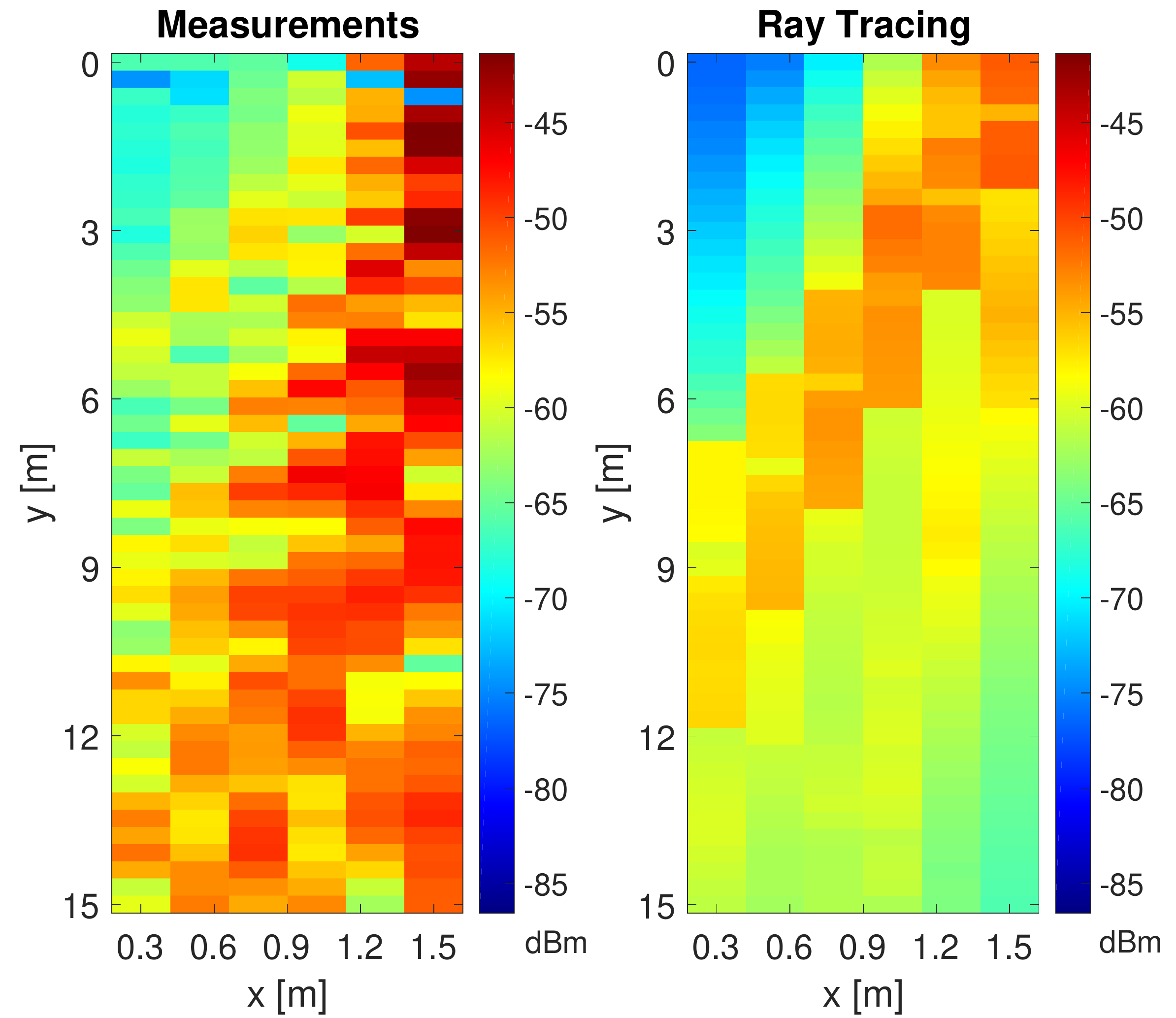}
	 \caption{}
     \end{subfigure}

\caption{Received power results for (a) no reflector,  obtained using (left) measurements, and (right) ray tracing simulations; (b) $12\times12$~in$^2$ flat square aluminum sheet at $\theta = 45^\circ$,  obtained using (left) measurements, and (right) ray tracing simulations; (c) $24\times24$~in$^2$ flat square aluminum sheet at $\theta = 45^\circ$,  obtained using (left) measurements, and (right) ray tracing simulations; (d) $33\times33$~in$^2$ flat square aluminum sheet at $\theta = 45^\circ$,  obtained using (left) measurements, and (right) ray tracing simulations.}\label{Fig:flat_combine}
\end{figure*}

\begin{figure*}[!h]
	\begin{subfigure}{0.5\textwidth}
	\centering
    \includegraphics[width=\columnwidth]{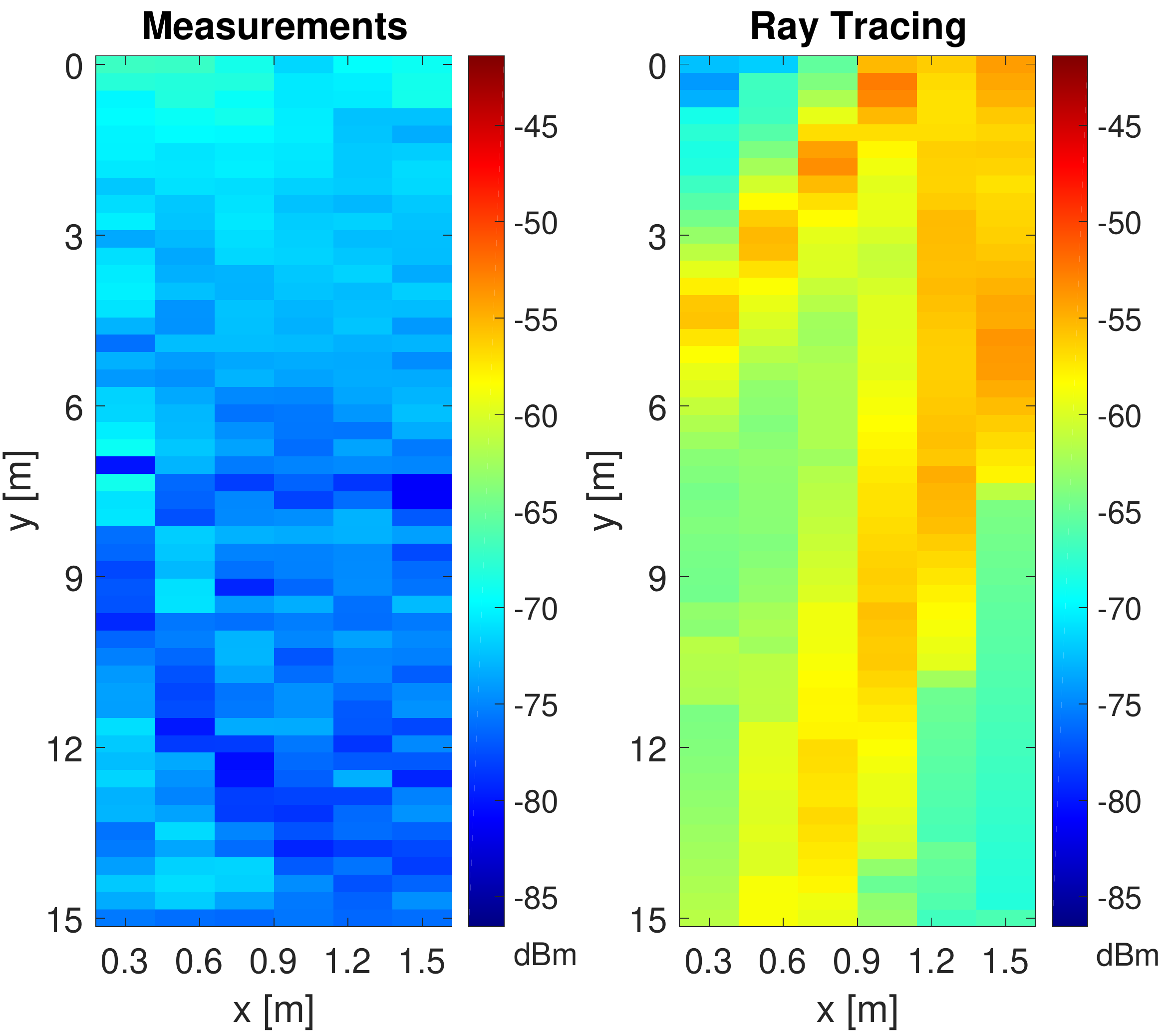}
	\caption{}
    \end{subfigure}       
    \begin{subfigure}{0.5\textwidth}
	\centering           
	\includegraphics[width=\columnwidth]{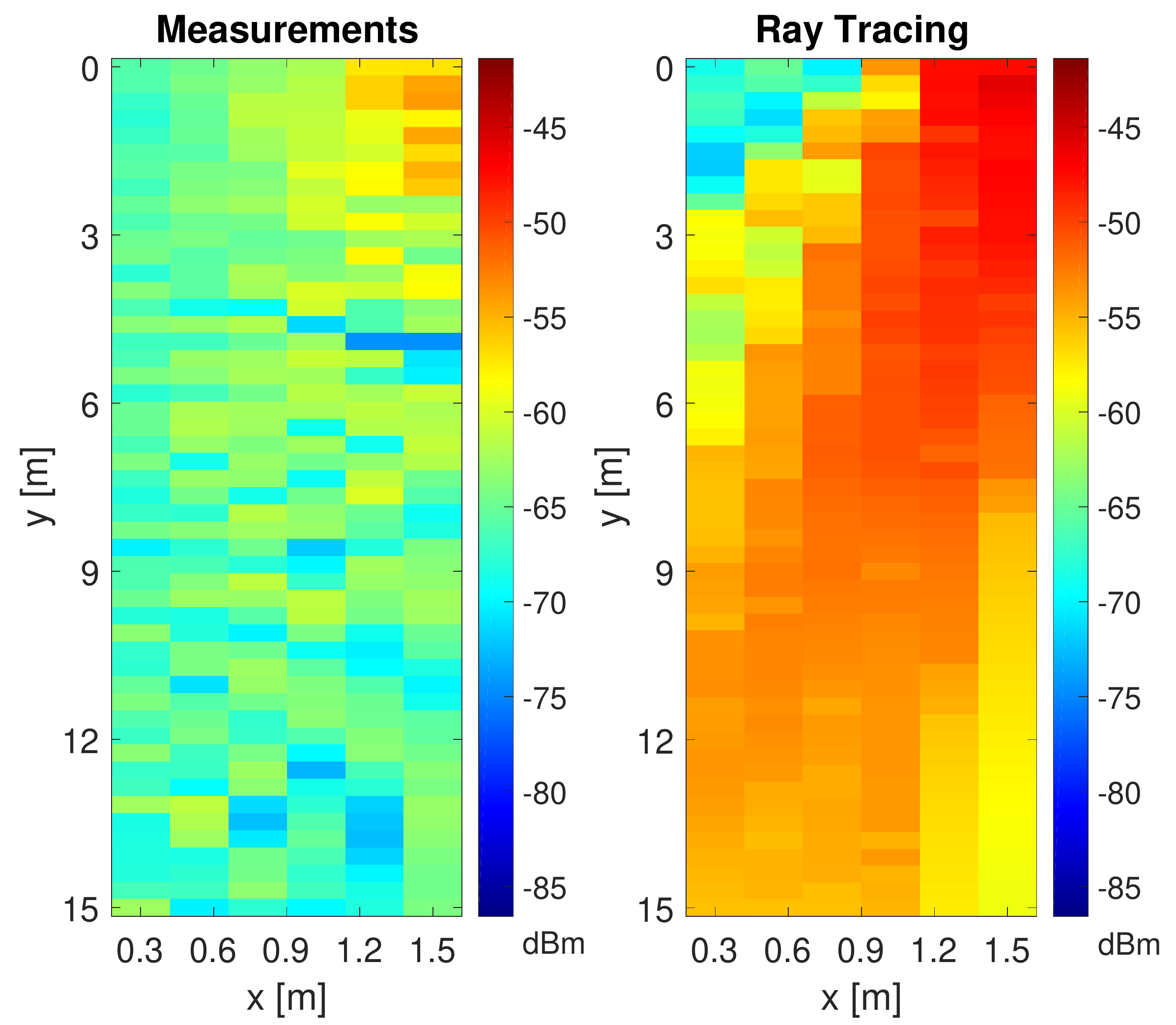}
	\caption{}
    \end{subfigure}    
   \caption{Received power results for (a) metallic sphere obtained using (left) measurements, and (right) ray tracing simulations; (b) metallic cylinder obtained using (left) measurements, and (right) ray tracing simulations.}\label{Fig:sphere_cylinder_combine} 
\end{figure*}

\begin{figure}
	\centering           
	\includegraphics[width=\columnwidth]{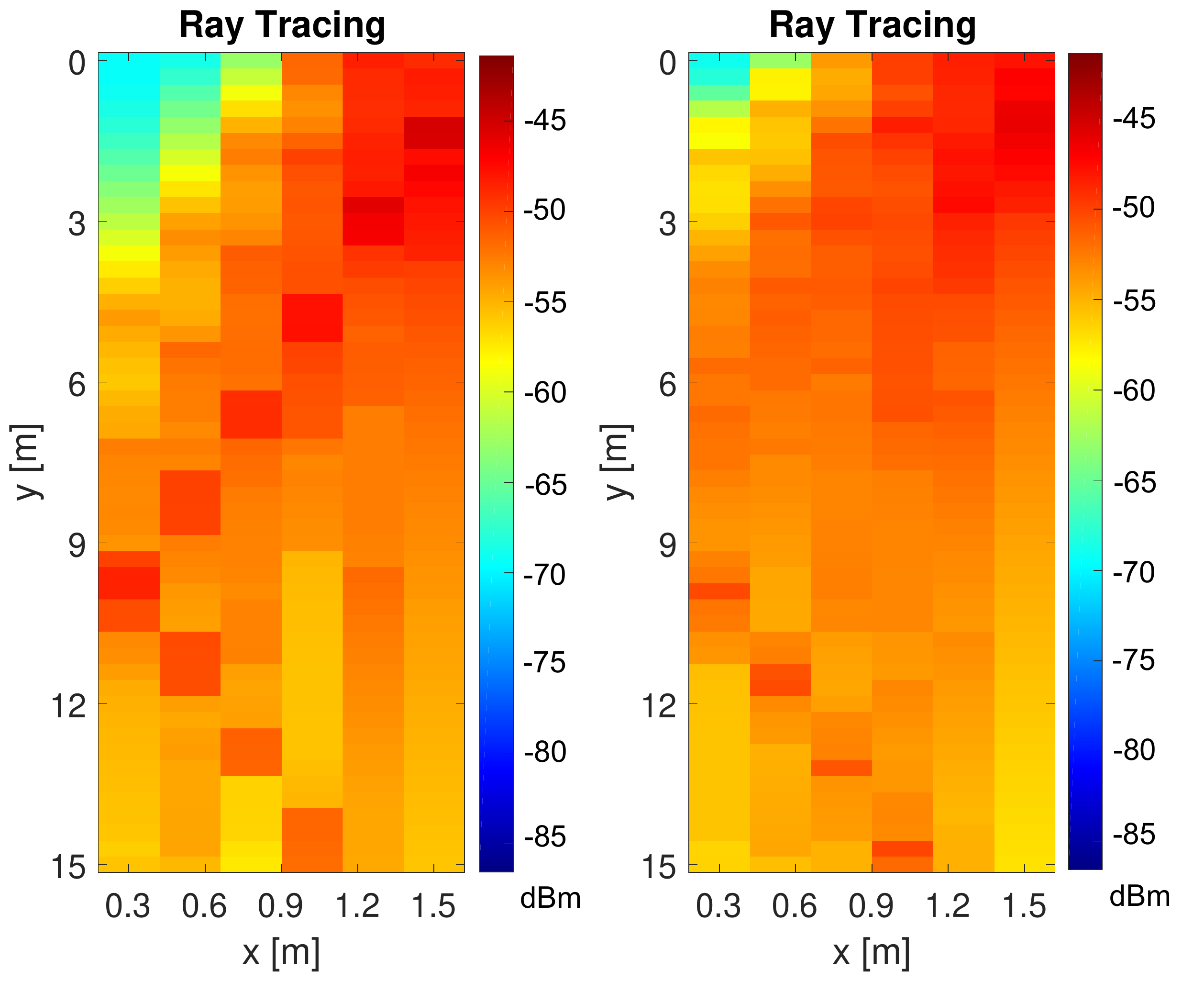}
	\caption{Received power results for metallic curved reflectors, (left) curve angle of $5$~degree, and (right) curve angle of $10$~degree, obtained using ray tracing simulations. } \label{Fig:curved}
\end{figure} 
    
\section{Indoor Empirical and Simulation Results}
In this section, empirical and simulation results are presented for the indoor NLOS measurements with and without metallic reflectors for the receiver grid as shown in~\ref{Fig:scenario}(a).

\subsection{Coverage with No Reflector}
In measurements shown in Fig.~\ref{Fig:flat_combine}(a) where no reflector is used, we observe slightly higher received power at the top left corner of the receiver grid mostly due to diffraction at the edge of the corridor wall. In case of simulations, we observe some reflections from the wall opposite to the transmitter; however, the overall received power in this case is less than the measurements.  

\subsection{Coverage with Square Metal Reflectors}
The flat reflectors are oriented at $45^\circ$ in the azimuth plane as shown in Fig.~\ref{Fig:scenario}(a) for all the measurements. 
For the $12\times12$~in$^2$ reflector shown in Fig.~\ref{Fig:flat_combine}(b), it can be observed that we have a directional coverage spreading with the distance along the y-grid. The reflected rays that are perpendicular to the incoming rays create a strip of dominant coverage area starting from the top right portion of the receiver grid. The width of this dominant coverage area is proportional to the width of the reflector, as discussed in Section~\ref{Section:Energy_Steering}. Furthermore, rays that are incident at different angles on the reflector gets reflected at corresponding angles. 

The $24\times24$~in$^2$ reflector measurement results are shown in Fig.~\ref{Fig:flat_combine}(c). Similar to $12\times12$~in$^2$ case, we observe a solid strip of dominant coverage area with a width proportional to the size of the reflector. Received power is distributed similarly across the receiver grid with a better coverage as compared to $12\times12$~in$^2$ case. 

The $33\times33$~in$^2$ reflector case is shown in Fig.~\ref{Fig:flat_combine}(d). It can be observed that we do not obtain additional benefit from the size of the reflector compared to $24\times24$~in$^2$ case. This can be explained due to the directional nature of the transmission. The surface area of the incident beam approximated to a rectangle will have higher intensity at the center as compared to the edges. Therefore, when using $33\times33$~in$^2$ reflector, we may cover more area of the incident beam. However, since the outer areas of the reflector receives lower energy as compared to the center, we observe no significant increase in the received power as compared to $24\times24$~in$^2$ reflector~(See Section~\ref{Section:Energy_Steering}).



For all the flat reflector scenarios, we observe power distribution mostly on the right side of the receiver grid, whereas we observe outage at the top left corner of the receiver grid. This is due to the property of directional reflection for the flat reflectors. Three plausible solutions to provide coverage on the top left side of the receiver grid, can be; 1) By orienting the reflector at angle less than $45^\circ$ (but will result in reduced power on the right side of the grid); 2) Using a transmit beam having higher angular diversity; 3) Using outward curved reflectors e.g. cylinders that can distribute the energy more uniformly on the grid due to divergence phenomenon.  

\begin{figure*}[!t]
	\begin{subfigure}{0.5\textwidth}
	\centering
    \includegraphics[width=\textwidth]{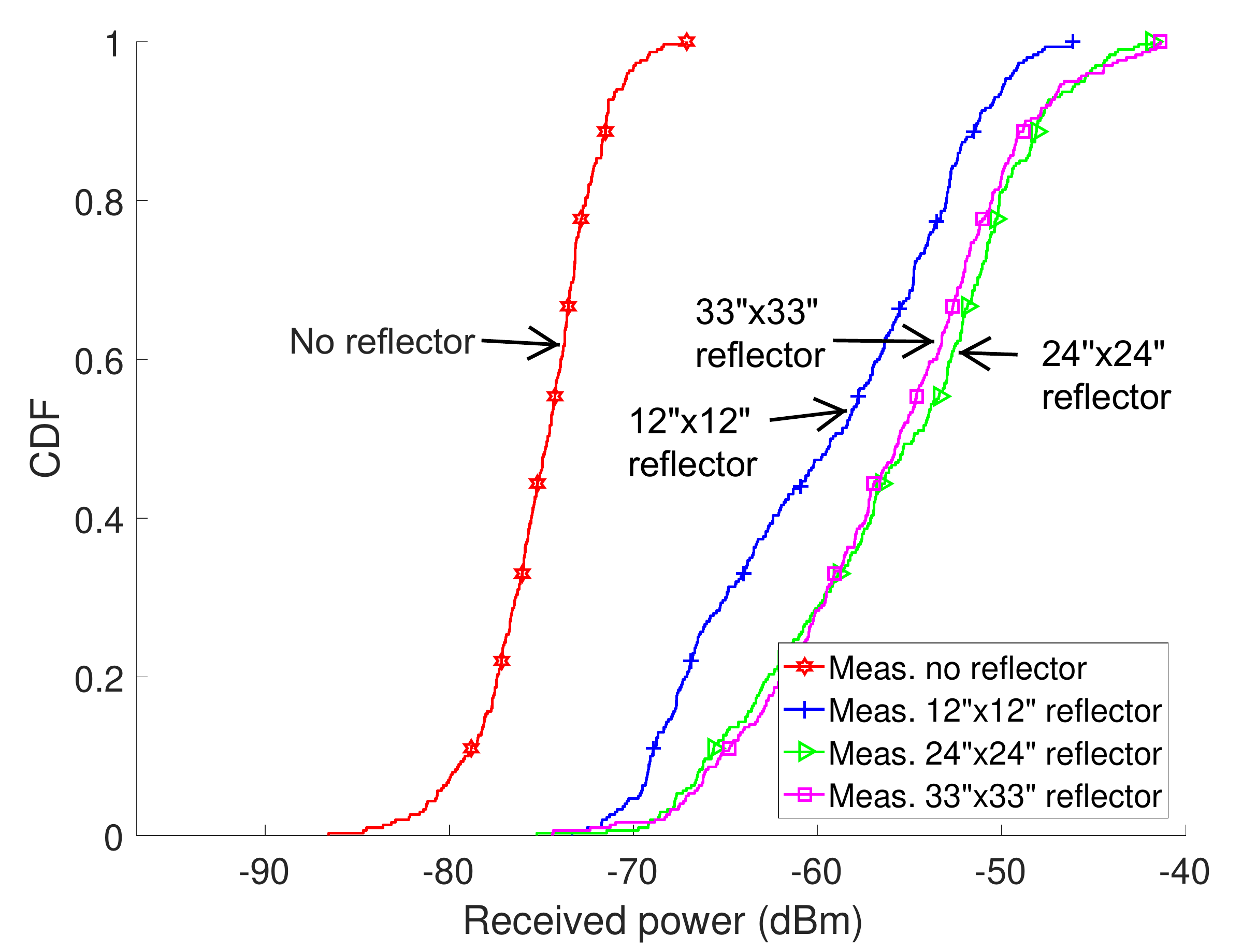}
    \caption{}
    \end{subfigure}	
 	\begin{subfigure}{0.5\textwidth}
    \centering
	\includegraphics[width=\textwidth]{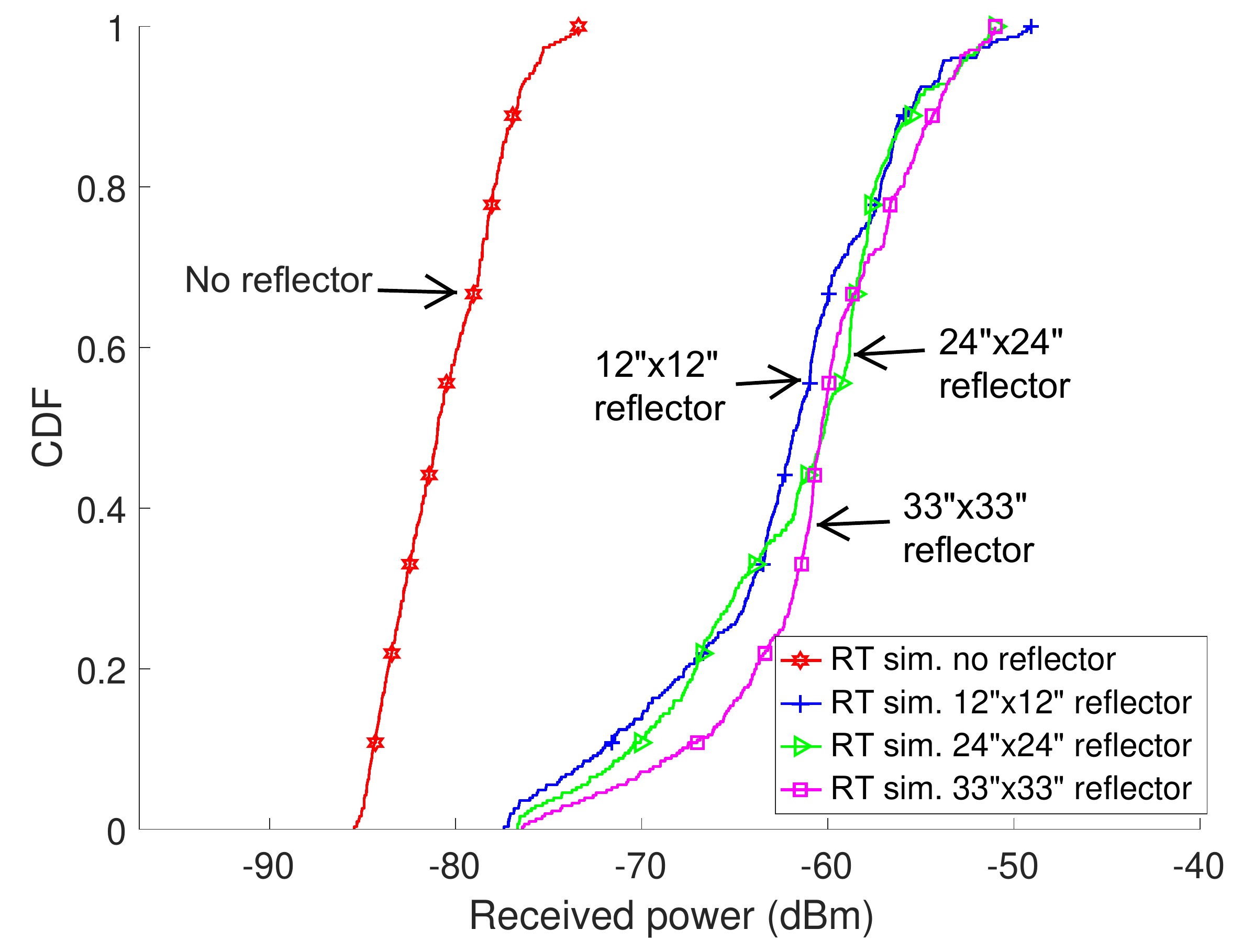}
	  \caption{}  
    \end{subfigure}     
	\begin{subfigure}{0.5\textwidth}
    \centering
	\includegraphics[width=\textwidth]{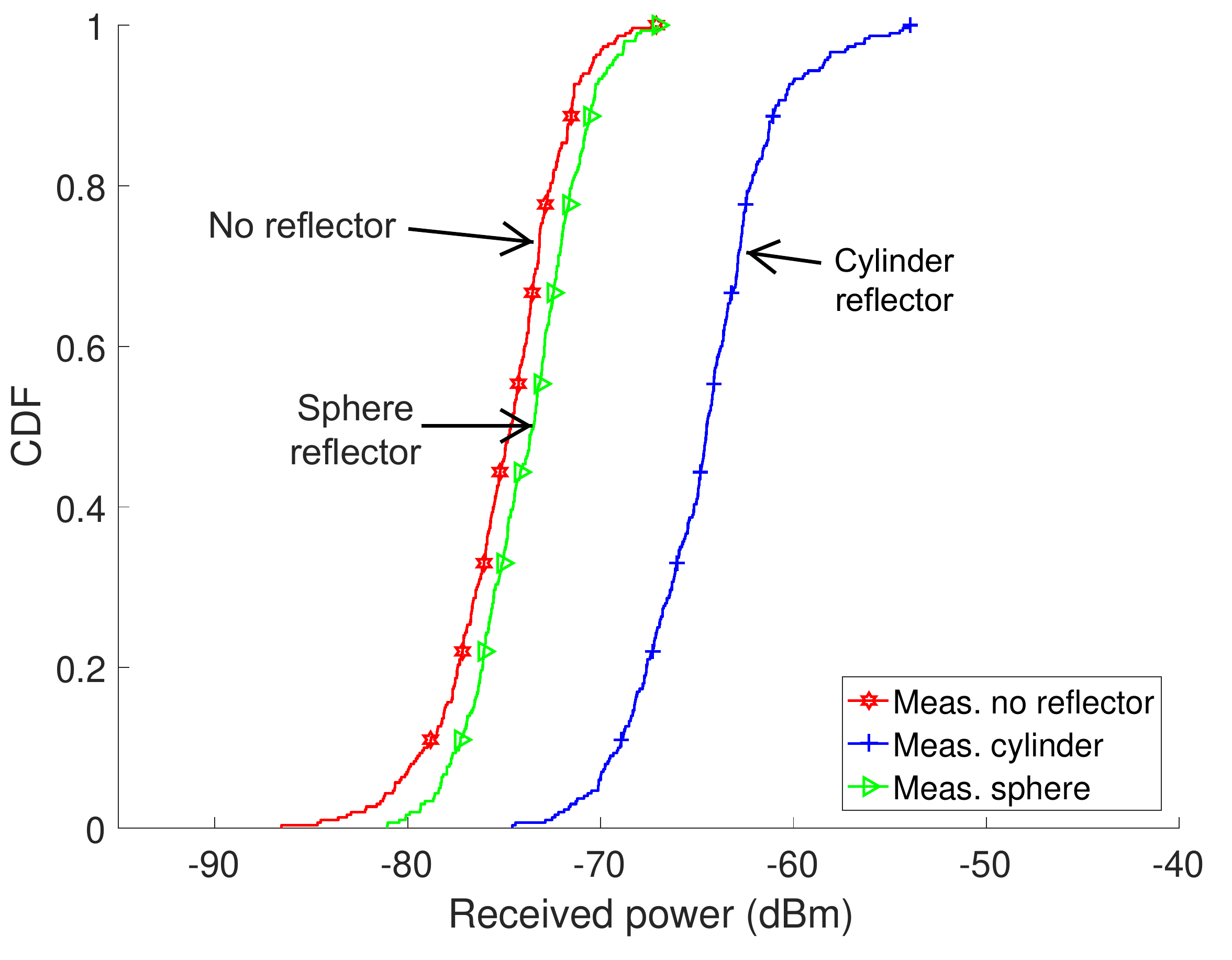}
	  \caption{}  
    \end{subfigure}    
    \begin{subfigure}{0.5\textwidth}
    \centering
	\includegraphics[width=\textwidth]{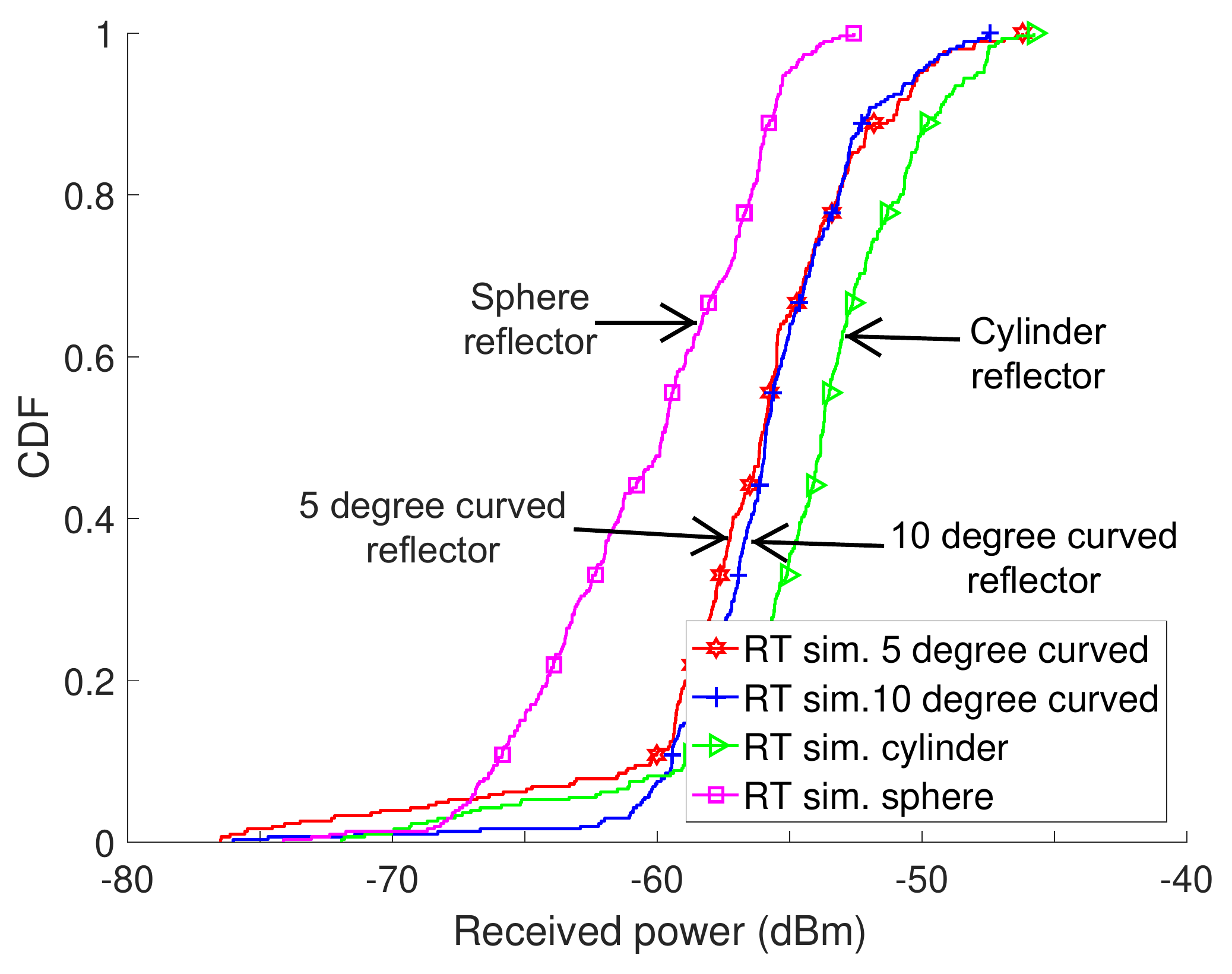}
	  \caption{}  
    \end{subfigure}
    
    \caption{CDF of received power for multiple scenarios. (a) no reflector, $12\times12$~in$^2$, $24\times24$~in$^2$, and $33\times33$~in$^2$ flat square sheet reflectors (measurements); (b) no reflector, $12\times12$~in$^2$, $24\times24$~in$^2$, and $33\times33$~in$^2$ flat square sheet reflectors (simulations); (c) no reflector, cylinder reflector, and sphere reflector (measurements); (d) cylinder, sphere, and curved reflectors (simulations).}\label{Fig:CDF_combine}
\end{figure*}

In Fig.~\ref{Fig:flat_combine} side by side comparison of measurements and RT simulations are possible. The distribution of the power on the receiver grid is similar for both the measurements and the simulations for the three different reflector sizes. For all the cases, we observe smaller received power for RT simulations when compared to the measurements. One reason is the presence of additional small scatterers in the environment and approximate diffuse scattering coefficients used in the simulations for real world materials. A small additional power gain is observed for $12\times12$~in$^2$ and $24\times24$~in$^2$ reflector from the cardboard which was not included in simulations.

\subsection{Coverage with Cylinder and Sphere Reflectors}
The measurement and simulation results for cylinder and sphere reflectors are shown in Fig.~\ref{Fig:sphere_cylinder_combine}. We observe that the sphere reflector does not help much to improve the performance. The cylinder reflector provides more uniform power distribution on the receiver grid as compared to flat reflectors. Although, both cylinder and sphere have curved shapes with equal surface areas, we observe higher received power for the cylinder. The reason for this can be explained due to high divergence of the incoming rays randomly in the surroundings from the top and bottom of the sphere. Similarly, for both the cylinder and sphere, we observe less power as compared to the flat reflector $24\times24$~in$^2$, though all three have approximately similar cross section area. The main reason for this behavior is due to small cross section area of the cylinder and sphere exposed to the incoming beam as compared to the flat reflector. Additionally, we observe high divergence of incoming rays in different directions from curved surfaces.

Simulation results in Fig.~\ref{Fig:sphere_cylinder_combine} exhibit larger received power as compared to measurements. This behavior can be explained due to limitations in the construction of the cylinder and sphere. As the cylinder was built by curving a metallic sheet, a slight difference in the curve angle of the cylinder can change the reflection of rays on the receiver grid. Similarly, sphere was constructed from a mirror ball with aluminum sheet wrapped on the top. The wrapping introduces wrinkles that result in additional random scattering of the incoming rays. 

\subsection{Coverage with Curved Reflectors}
Due to directional properties of the flat reflector, we expect most of the energy directed in a given direction on the receiver grid as shown in Fig.\ref{Fig:flat_combine}. However, an outage is observed on the receiver grid at the top left of the grid for these reflectors. This outage is minimized using cylinder and sphere reflector as shown in Fig.~\ref{Fig:sphere_cylinder_combine}. However, only limited area of the cylinder and sphere reflector is exposed to the incident rays. Therefore, in order to combine the properties of cylinder and sphere, we have built curved reflectors at different angles of curve in AutoCAD in order to uniformly reflect the incoming rays with maximum surface area exposed to the incoming rays as shown in Fig.~\ref{Fig:setup_reflectors}(e). The surface area of these curved reflectors with a height of $24$~in is approximately similar to the $24\times24$~in$^2$ flat, sphere and cylinder reflectors.

RT simulations are performed for curved reflectors with two curve angles of $5$~degree and $10$~degree, respectively. The received power results are shown in Fig.~\ref{Fig:curved}. We observe that $10$~degree curved reflector mostly solves the coverage problem we observe at the top left portion of the receiver grid. This is mostly due to higher divergence of the rays when incident on the curved surface. For $5$~degree curved reflector, we observe slightly higher outage than the $10$~degree reflector at the top left side of the receiver grid. In the case of smaller curve angle, the divergent rays do not reach the top left side of the receiver grid.


\subsection{CDF of Received Power with/without Reflector}
The power values across the whole receiver grid can be combined into a single cumulative distribution function (CDF) plot for each scenario. The CDF plots of received power over the whole receiver grid for flat reflectors and no reflector are shown in Fig.~\ref{Fig:CDF_combine}(a). As observed previously, the received power for the $12\times12$~in$^2$ reflector is smaller than the $24\times24$~in$^2$ and $33\times33$~in$^2$ reflectors, while the low power~(outage) areas are similar. The variance of the received power with different reflectors is higher when compared to no reflector scenario as expected. Another observation from Fig.~\ref{Fig:CDF_combine}(a) is that the received power varies in the range [-$75$,-$40$]~dBm, while for no reflector case it is [-$85$,-$70$]~dBm. This behavior can be related to directional propagation in mmWave bands. In particular, highly directional scattering (reflection) results in power increase in some regions less and in some others more. In Fig.~\ref{Fig:CDF_combine}(b), the CDFs of received power from the simulations are shown corresponding to same reflector scenarios as in Fig. 6(a). Results show that simulations match reasonably with measurements. We obtain a median gain of around $20$~dB for the $24\times24$~in$^2$ and $33\times33$~in$^2$ reflector scenarios as compared to the no reflector case. 

The CDF of received power for cylinder and sphere reflectors from measurements and simulations are shown in Fig.~\ref{Fig:CDF_combine}(c) and Fig.~\ref{Fig:CDF_combine}(d), respectively. Cylinder reflector exhibits  higher received power in the measurements compared to the sphere. Whereas in simulations, we observe high received power for both cylinder and sphere. The CDF of the received power for the curved reflectors with curve angles of $5$~degree and $10$~degree obtained using RT simulations are also shown in Fig.~\ref{Fig:CDF_combine}(d). In particular, the 10 degree curved reflector performs the best when the 10th percentile worst case user region is considered, while cylindrical reflector is the best for users having higher received power than the 10th percentile received power.  

\section{Conclusions}\label{Section:Concluding Remarks}
In this work, channel measurements at $28$~GHz are carried out in an NLOS indoor scenario. Passive metallic sheet reflectors of different shapes and sizes are used to enhance the received power, yielding a better signal coverage in the NLOS region. It is observed that increasing the size of reflector beyond a certain value at a given distance (of transmitter from the reflector) may not result in significant increase in the reflection power. Additionally, the coverage area enhancement using reflectors is found to be dependent on the radiation pattern of the antenna, and the size of the reflector at a given distance of the transmitter from the reflector. Maximum power is obtained at an azimuth angle of $45^{\circ}$ for $24\times24$~in$^2$ and $33\times33$~in$^2$ sized reflectors, where we observe a median gain of $20$dB when compared to no reflector case. RT simulations suggested that the outage problem observed in top left portion of the receiver grid may be solved using curved reflector. The measurement results were compared with RT simulations that provides a close agreement. 

\section*{Acknowledgement}
This work has been supported in part by NASA under the Federal Award ID number
NNX17AJ94A and by DOCOMO Innovations, Inc.


\bibliographystyle{IEEEtran}

\begin{thebibliography}{10}
\providecommand{\url}[1]{#1}
\csname url@samestyle\endcsname
\providecommand{\newblock}{\relax}
\providecommand{\bibinfo}[2]{#2}
\providecommand{\BIBentrySTDinterwordspacing}{\spaceskip=0pt\relax}
\providecommand{\BIBentryALTinterwordstretchfactor}{4}
\providecommand{\BIBentryALTinterwordspacing}{\spaceskip=\fontdimen2\font plus
\BIBentryALTinterwordstretchfactor\fontdimen3\font minus
  \fontdimen4\font\relax}
\providecommand{\BIBforeignlanguage}[2]{{%
\expandafter\ifx\csname l@#1\endcsname\relax
\typeout{** WARNING: IEEEtran.bst: No hyphenation pattern has been}%
\typeout{** loaded for the language `#1'. Using the pattern for}%
\typeout{** the default language instead.}%
\else
\language=\csname l@#1\endcsname
\fi
#2}}
\providecommand{\BIBdecl}{\relax}
\BIBdecl

\bibitem{FCC_28G}
\BIBentryALTinterwordspacing
{Federal Communications Commission}, ``{FCC} rules for next generation wireless
  technologies,'' accessed: 1-12-2018. [Online]. Available:
  \url{https://www.fcc.gov/document/fcc-adopts-rules-facilitate-next-generation-wireless-technologies}
\BIBentrySTDinterwordspacing

\bibitem{Light_EM}
M.~Kerker, \emph{The scattering of light and other electromagnetic
  radiation}.\hskip 1em plus 0.5em minus 0.4em\relax Elsevier, 2016.

\bibitem{reflection}
S.~N. Ghosh, \emph{Electromagnetic theory and wave propagation}.\hskip 1em plus
  0.5em minus 0.4em\relax CRC Press, 2002.

\bibitem{NASA_refl}
C.~C. Cutler, ``Passive repeaters for satellite communication systems,'' Feb.~9
  1965, uS Patent 3,169,245.

\bibitem{Literature4}
J.~L. Ryerson, ``Passive satellite communication,'' \emph{Proc. of the IRE},
  vol.~48, no.~4, pp. 613--619, April 1960.

\bibitem{Literature5}
Y.~E. Stahler, ``Corner reflectors as elements passive communication
  satellites,'' \emph{IEEE Trans. Aerospace}, vol.~1, no.~2, pp. 161--172, Aug.
  1963.

\bibitem{Microwave_refl}
Y.~Huang, N.~Yi, and X.~Zhu, ``Investigation of using passive repeaters for
  indoor radio coverage improvement,'' in \emph{IEEE Ant. Propag. Society
  Sympos.}, vol.~2, June 2004, pp. 1623--1626 Vol.2.

\bibitem{Literature6}
J.~L. D. L.~T. Barreiro and F.~L.~E. Azpiroz, ``Passive reflector for a mobile
  communication device,'' Aug. 2006, uS Patent 7,084,819.

\bibitem{lit_60GHz_indoor}
Z.~Genc, U.~H. Rizvi, E.~Onur, and I.~Niemegeers, ``Robust 60 {GHz} indoor
  connectivity: is it possible with reflections?'' in \emph{IEEE Vehic.
  Technol. Conf. (VTC)}, Spring, 2010, pp. 1--5.

\bibitem{Literature3}
Z.~Peng, L.~Li, M.~Wang, Z.~Zhang, Q.~Liu, Y.~Liu, and R.~Liu, ``An effective
  coverage scheme with passive-reflectors for urban millimeter-wave
  communication,'' \emph{IEEE Ant. Wireless Propag. Lett.}, vol.~15, pp.
  398--401, 2016.

\bibitem{Literature1}
M.~Heino, C.~Icheln, and K.~Haneda, ``Reflector design to mitigate finger
  effect on 60 {GHz} user devices,'' in \emph{European Conference on Antennas
  and Propagation (EUCAP)}, March 2017, pp. 151--155.

\bibitem{Literature2}
A.~A. Goulianos, T.~H. Barratt, W.~Yuan, S.~Zhang, M.~A. Beach, A.~R. Nix,
  E.~Mellios, P.~Cain, M.~Rumney, and T.~Masson, ``Time-domain sounder
  validation and reflectivity measurements for {mm-Wave} applications,'' in
  \emph{Proc. IEEE Wireless Commun. Netw. Conf. (WCNC)}, April 2016, pp. 1--5.

\bibitem{Antenna_approx}
U.~A. Department, \emph{Manuals Combined: Electronic Warfare and Radar
  Systems}.\hskip 1em plus 0.5em minus 0.4em\relax Naval Air Warfare Center
  Weapons Division Point Mugu, CA, 2013.

\bibitem{NImmwave}
\BIBentryALTinterwordspacing
{National Instruments}, ``{mmWave} transceiver system,'' accessed: 1-15-2018.
  [Online]. Available: \url{http://www.ni.com/sdr/mmwave/}
\BIBentrySTDinterwordspacing

\bibitem{Horn_antenna_sage}
\BIBentryALTinterwordspacing
{Sage Millimeter, Inc }, ``{WR}-34 pyramidal horn antenna,'' accessed:
  1-10-2018. [Online]. Available:
  \url{https://www.sagemillimeter.com/content/datasheets/SAR-1725-34KF-E2.pdf}
\BIBentrySTDinterwordspacing

\end{thebibliography}





\end{document}